# Quantum Optical Metrology — The Lowdown on High-N00N States

Jonathan P. Dowling

Hearne Institute for Theoretical Physics and Department of Physics and Astronomy
Louisiana State University, 202 Nicholson Hall, Baton Rouge, Louisiana 70803 USA
Email: jdowling@lsu.edu

Quantum states of light, such as squeezed states or entangled states, can be used to make measurements (metrology), produce images, and sense objects with a precision that far exceeds what is possible classically, and also exceeds what was once thought to be possible quantum mechanically. The primary idea is to exploit quantum effects to beat the shot-noise limit in metrology and the Rayleigh diffraction limit in imaging and sensing. Quantum optical metrology has received a boost in recent years with an influx of ideas from the rapidly evolving field of optical quantum information processing. Both areas of research exploit the creation and manipulation of quantum entangled states of light. We will review some of the recent theoretical and experimental advances in this exciting new field of quantum optical metrology, focusing on examples that exploit a particular two-mode entangled photon state — the High-N00N state.

# 1. Introduction

In 1887 Albert Michelson and Edward Morley pioneered the precision use of an optical interferometer to measure the speed of light so accurately that they disproved the existence of the luminiferous ether (thought to be the medium in which light waves propagated) [1]. This feat paved the way for Albert Einstein's Special Theory of Relativity [2]. The Michelson interferometer (MI), depicted in figure 1, remains in use still today as a test bed for the theory of relativity, but this time for Einstein's General Theory. The interferometer is a giant antenna that searches the Heavens for gravity waves, which are predicted by the theory of Einstein, but which have never been directly observed. The Laser Interferometer Gravitational Wave Observatory (LIGO) is composed of two Michelson interferometers located in Louisiana and Washington State [3]. With a circulating laser power of about 100 kilowatts, and interferometer arms four kilometers long, these L-shaped machines are capable of measuring length displacements between the arms on the order of an attometer ($10^{-18}$ meters) — a thousand times smaller than the diameter of a proton! Surprisingly, the ultimate sensitivity of this gigantic, classical-looking device is limited by the quantum mechanical fluctuations of the circulating photon field and, ultimately, the electromagnetic fluctuations of the vacuum — that is, empty space itself. The newly emergent fields of quantum metrology, imaging, and sensing seek to exploit some of the same subtle effects exploited in quantum information processing, particularly quantum entanglement, to push the sensitivity of LIGO and all sorts of related interferometers to the ultimate quantum limit of resolution. Quantum optical metrology is a new field that specifically exploits these quantum effects to increase the signal-to-noise ratio in an array of sensors from LIGO-like interferometers to synchronized atomic clocks. Quantum imaging exploits similar quantum ideas to, for example, beat the Rayleigh diffraction limit in resolution of an imaging system, such as used in optical lithography. Quantum sensing is a new area of quantum technology that seeks to exploit the advances in quantum metrology and imaging in practical remote sensors, such as laser Light Detection and Ranging (LIDAR) systems, with scientific, commercial, and defense applications.

To understand the role of quantum mechanics in optical interferometers, we first consider the prototypical Mach-Zehnder interferometer (MZI) as shown in figure 2. The MZI is an unfolded MI, in that the circulating light makes one pass through two separated beam splitters (BS), instead of two passes through one BS, as was the case in the Michelson machine of figure 1. The two interferometers are mathematically equivalent. In the MI it is a bit easier to align the laser beams and it is preferred in large interferometer applications such as LIGO, but the MZI is a bit easier to draw and analyze mathematically and is preferred by theorists and in small experimental test beds. The results extracted from the MZI apply as well to the MI.

So let us set our task to measure, as accurately as possible, the path-length difference between the two branches (or arms) of the interferometer using monochromatic light of wavelength $\lambda$. (In LIGO the wavelength is about one micron.) In the standard approach, used in LIGO, a laser beam is launched into the first fifty-fifty beam splitter (50-50 BS) on the left in port A, bounced off of the two mirrors in the middle, and recombined at the second 50-50 BS on the right. Light then emerges from the top and bottom ports, C and D, of the second beam splitter and is then made incident on two photodetectors $\mathcal{C}$ and $\mathcal{D}$, as shown. Typically the intensities in each port, $I_C$ and $I_D$, are measured at each detector and the result is combined to yield the difference intensity, $I = I_C - I_D$, which we shall call the signal. To indicate the phase induced by the path difference between the upper and lower branches, we place an icon for a phase shifter $\varphi$ that, in this example, has the value $\varphi = kx$, where $x$ is the path difference between the two arms, which is the quantity to be measured. The wave number, $k = 2\pi/\lambda$, is a known constant, given the optical wavelength $\lambda$. (For a typical laser the spread or uncertainty in the wavelength is very small.) The idea is to use the light beam itself as ruler, with tick marks spaced by units of $\lambda$, in order to measure the path-length difference $x$. This can be done by first balancing the interferometer, that is by making $x = 0$. In this balanced case the light travels exactly the same distance via the upper and the lower branch.

We adopt the convention that the light field always picks up a $\pi/2$ phase shift upon reflection off of a mirror or off of a BS, and also no phase shift upon transmission through a BS [4]. Under these circumstances, the two light fields emerging from the second BS out the upper port C are precisely $\pi$ out of phase with each other, and hence completely cancel out due to destructive interference. (This is called the dark port.) Consequently the two light fields recombine completely in phase as they emerge from the lower port D and add up due to constructive interference. (This is called the bright port.) Hence for a balanced MZI all of the energy that comes in port A emerges out of port D and none out port C. Clearly, any change in the path difference $x$ away from the $x = 0$ balanced condition will cause light to appear in the formerly dark port, and in this way we can measure $x$ by simply measuring intensities at the detectors. This is in fact what LIGO does. How precise a measurement of the path difference $x$ can we make?

It is straightforward to show, using classical optical theory, that if the light intensity incident on port A is $I_A$ then in terms of the phase shift $\varphi$ the output-port intensities may be written as [4],

$$I_C = I_A \sin^2(\varphi/2), \tag{1a}$$

$$I_D = I_A \cos^2(\varphi/2). \tag{1b}$$

Energy is conserved as, from a simple trigonometry identity, $I_C + I_D = 1$, for all values of $\varphi$. It is typical for the analyzer in figure 2 to compute the difference intensity $M = I_D - I_C$ (where $M$ stands for "minus"). Again, using simple trigonometry, we can write,

$$M(\varphi) \equiv I_D - I_C = I_A \cos(\varphi). \tag{2}$$

We plot $M$ in figure 3 as a function of $\varphi$, and we can see it varies periodically. Since $\varphi/2 = kx/2 = \pi x/\lambda$, from properties of sine and cosine, we have that $I_C = 0$ and $I_D = I_A$ whenever $x/\lambda = 0, 1, 2, 3, \ldots$. Hence our ruler is the light wave itself and the tick marks are spaced the wavelength $\lambda$ apart. We may start with a balanced interferometer with equal arm lengths, *x = 0* (and $M = I_A$) and then slowly move the upper mirror upwards increasing *x*. As we do we will break the balance and begin to see light emerging from the formally dark port C (*M* decreases in the plot). At the point $\varphi = \pi/2$, when $I_C = I_D$, then *M = 0*. Eventually we will see port C attain maximum brightness and port D will go dark ($M = -I_A$). As we continue the mirror displacement this process will reverse, as sine and cosine are periodic, and finally port C will go dark again (*M* is maximum again with $M = I_A$). At this point we can stop moving the upper mirror and we are assured that now the path difference *x* has gone from *0* to $\lambda$. If we take $\lambda = 1.0$ µm (about what is used in LIGO) then it would seem we have a machine capable of measuring distances to an accuracy of about $\lambda = 1.0$ µm. This is consistent with the Rayleigh diffraction limit, typically invoked in classical optics, and so everything looks hunky dory. But this is not the end of the story. Recall that I mentioned above that gravity waves are expected to cause displacements in the LIGO mirrors by $10^{-18}$ meters. A micron is only $10^{-6}$ meters, and so we have come up twelve-orders-of magnitude too short of our goal for measuring gravity waves! We can actually do much, much, much better than one micron, by exploiting the quantum mechanical nature of light.

Let us then now consider a different strategy to estimate the precision of the device. Let us balance the interferometer such that we start at the point $\varphi = \pi/2$ when $I_C = I_D$ and hence *M = 0* in figure 3. Note this is where the curve crosses the horizontal axis and the slope of the *M*-curve is steepest. Let us continue with the analysis. If we call the horizontal displacement change $\Delta\varphi$, then we can see this is related to the vertical intensity change $\Delta M$. For small changes we may approximate this relation using differentials, that is, $\Delta M = I_A \Delta \varphi$, or,

$$\frac{\Delta M}{\Delta \varphi} = \frac{\partial M}{\partial \varphi} = I_A \sin(\varphi), \tag{3}$$

which may be written,

$$\Delta \varphi = \frac{\Delta M}{\partial M / \partial \varphi} = \frac{\Delta M}{I_A \sin(\varphi)}. \tag{4}$$

The quantity $\partial M / \partial \varphi$ is the slope of the curve, which is largest at the crossing point, implying our minimum detectible phase $\Delta \varphi$ is smallest there, via Eq. (4). At the crossing point $\varphi = \pi/2$ and $\sin(\pi/2) = 1$, and so this relation would seem to indicate that *if* we can measure the intensity displacement $\Delta M$ with infinite precision ($\Delta M = 0$), we can measure the phase (and hence distance) with infinite precision ($\Delta \varphi = 0$). Hence it would appear, with our new scheme, that we could detect any gravity wave no matter how small its amplitude, or how far away its source! This is too good to be true, of course. So we now have two different, classical, estimates putting our length measuring precision somewhere between zero and one micron. The truth lays somewhere in between these two extremes.

The problem is that the simple classical arguments we used above do not take into account the effects of quantum mechanics. Specifically it does not take into account the fact that the intensity of the light field is not a constant, which can be measured with infinite precision, but that it fluctuates about some average value, and those fluctuations have their origin in the vacuum fluctuations of the quantized electromagnetic field. According to quantum mechanics, optical intensity can never be measured with infinite precision. Hence the uncertainty in the red curve of figure 3, always has some finite value, indicated by the box of height $\Delta M$. The intensity displacement $M$ can never be measured with infinite precision and has a fundamental uncertainty $\Delta M$, and therefore the consequent phase $\varphi$ will always have its related uncertainty $\Delta \varphi$, which is the width of the box. (This immediately implies a minimum detectable displacement $\Delta x = \lambda \Delta \varphi$.) These fundamental quantum intensity fluctuations suggest that there is a Heisenberg uncertainty principal at work, which in our example implies that the intensity $I$ and the phase $\varphi$ cannot both be measured with infinite precision simultaneously. There is indeed such an uncertainty relation, as we shall see next.

For a quantum analysis of this phenomenon, we introduce the mean number of photons in the laser field as the dimensionless quantity $n$, and note that the intensity $I$ is then proportional to $n$ for a steady-state system. If we denote the fluctuation in the phase as $\Delta \varphi$ and that in the intensity as $\Delta n$, we can then write down the Heisenberg number-phase uncertainty relation as

$$\Delta n \Delta \varphi \geq 1 \tag{5}$$

Paul Adrien Maurice Dirac first proposed this uncertainty relationship between photon number and phase [5]. It is closely related to the better know energy-time uncertainty principal $\Delta E \Delta t \geq \hbar$, where $\Delta E$ is the uncertainty in the energy, $\Delta t$ is the uncertainty in the time, and $\hbar$ is Dirac's constant (Planck's constant divided by $2\pi$). Starting with the energy-time principal, we can give a heuristic derivation of the number-phase formula. For a standing, monochromatic, electromagnetic wave we have $E = \hbar n \omega$, where $\omega$ is the frequency ($\omega = k/c$ where $c$ is the speed of light). This is just the energy per photon multiplied by the mean number of photons. Since there is no propagation for a standing wave we have $\varphi = \omega t$ as the accumulated phase at any point. Approximating both of these expressions with differentials, holding everything except $E$ and $t$ constant, gives $\Delta E = \hbar \Delta n \omega$ and $\Delta \varphi = \omega \Delta t$. Inserting these two expressions into the energy-time uncertainty relation yields the number-phase relation, Eq. (5).

For a laser beam, such as used in LIGO, the quantum light field is well approximated by a coherent state, denoted $|\alpha\rangle$, where the complex number $\alpha = |\alpha| e^{i\varphi}$ is proportional to the electric field amplitude $E$ such that $|\alpha|^2 = n$, the latter of which we recall is the dimensionless field intensity. (This is the dimensionless quantum version of the classical relation $|E|^2 = I$.) The fluctuations are typically represented in a phasor diagram as shown in figure 4. Here the phase is the polar angle $\varphi$ is measured counter-clockwise off

the horizontal axis. The radius from the origin to the center of the coherent-state disk is $R = |\alpha|^2 = n$. The diameter of the disk $d$ is on the order of $d = \Delta n = \sqrt{n}$. From simple geometry, we can then approximate $d = R\, \Delta\varphi$, where $\Delta\varphi$ is the uncertainty or fluctuation in the angular $\varphi$ direction. Combining all this we arrive at the fundamental relationships between number (intensity) and phase uncertainty for a coherent-state laser beam,

$$\Delta n\, \Delta\varphi = 1, \tag{6a}$$

$$\Delta n = \sqrt{n}, \tag{6b}$$

$$\Delta\varphi_{SNL} = \frac{1}{\Delta n} = \frac{1}{\sqrt{n}}. \tag{6c}$$

The first relation, Eq. (6a), tells us that we have equality in Eq. (5), that is a coherent state is a minimum uncertainty state (MUS). Such a state saturates the Heisenberg number-phase uncertainty relation with equality. This is the best you can do according to the laws of quantum mechanics. The second relation, Eq. (6b), describes the fact that the number fluctuations are Poissonian with a mean of $n$ and a deviation of $\Delta n = \sqrt{n}$, a well-known property of the Poisson distribution and the consequent number statistics for coherent-state laser beams [4]. Putting back the dimensions we arrive at,

$$\Delta\varphi_{SNL} \propto \frac{1}{\sqrt{I}}, \tag{7}$$

which is called the shot-noise limit (SNL). (The term "shot noise" comes from the idea that the photon-number fluctuations arise from the scatter in arrival times of the photons at the beam splitter, much like buckshot from a shotgun ricocheting off a metal plate.) We can also import the SNL into our classical analysis above. Consider Eq. (4), where we now take $I_A = n$, $\Delta M = \sqrt{n}$, and $\varphi = \pi/2$. We again recover Eq. (6c) for the phase uncertainty. Hence quantum mechanics puts a quantitative limit on the uncertainty of the optical intensity, and that intensity reflects itself in a consequent quantitative uncertainty of the phase measurement.

So now we can see, by exploiting quantum mechanics, we can do much better than the one-micron accuracy that Lord Rayleigh might have expected, by simply increasing the power or intensity of the light field. However we can never have perfect precision, as our second naïve argument indicated, because we would have to have infinite intensities in the laser beam. If we recall that $\varphi = kx$ then approximating with differentials we have $\Delta\varphi = k\Delta x$ (since $k = 2\pi/\lambda$ is a constant) and we obtain the minimum detectable distance as,

$$\Delta x_{SNL} = \frac{\lambda}{\sqrt{n}} \propto \frac{\lambda}{\sqrt{I}}, \tag{8}$$

where $\bar{\lambda} = \lambda / (2\pi)$. We can see from this expression that we can do much, much, much better than a Rayleigh length resolution on the order of $\lambda$ by just cranking up the laser power. The LIGO observatories currently have a circulating power in the interferometer on the order of 100 kilowatts. This corresponds to a mean photon number of about $n = 10^{24}$ photons in the device at a time. Hence we have an enhancement factor in resolution of $\sqrt{n} = 10^{12}$ over the Rayleigh limit. Therefore the minimal detectable displacement of the mirrors is given by Eq. (8) as $\Delta x = (10^{-6} \, \mu m) (10^{-12}) = 10^{-18}$ meters, which is right about where the LIGO folks expect to start seeing gravity waves. (The effect of a passing gravity wave is to cause displacement between the two arms of the MI by about this amount.) In fact, the LIGO interferometer is already constrained by this shot-noise limit over an appreciable range of interferometer oscillation frequencies where gravity waves are expected. (None have been seen yet.) While a length measurement accurate to one part in $10^{20}$ may seem astounding, and in fact LIGO is really one of the most sensitive measuring devices ever built, what is really surprising is that our bag of tricks is not exhausted yet. It is possible to do even better still with a bit more elaborate quantum trickery.

## 2. Squeezed States of Light and the Heisenberg Limit

In 1981 Carlton Caves first proposed the idea of using non-classical states of light — the so-called squeezed states — to improve the sensitivity of optical interferometers to even below the shot-noise limit [6]. This notion came as somewhat of a surprise to the interferometer community, as it was thought at the time that the shot-noise limit was the ultimate limit on sensitivity as imposed by quantum mechanics. I like to tell my students "There is no such thing as classical mechanics!" What I mean by this is that all physical systems are quantum mechanical in nature; the only question is if the classical theory is a sufficiently accurate approximation to reality so that it can be used instead of quantum theory. To use or not to use quantum theory? — That is the question.

For example, one might guess for a four-kilometer long optical interferometer with a circulating power of 100 kilowatts, that a description based on classical theory would be just fine. This guess would be wrong, as we have shown in the previous section. The classical argument implies a sensitivity of microns but the simple quantum argument buys us down to attometers. It is because LIGO is so precise, and because all classical sources of noise have been eliminated, that we are forced at the end to deal with the quantum nature of the beast itself. Nevertheless, quantum opticians call the coherent state of laser light classical, in spite of the fact that it exhibits fluctuations which are quantum mechanical in origin and that limit the sensitivity of LIGO. A better description would be to say that a coherent state of light is as classical as you can get. So there is still some maneuvering room here in the quantum game.

In classical electromagnetism we can also represent a monochromatic plane wave on the phasor diagram of figure 4 — but instead of a disk the classical field is depicted as a point. The radial vector to the point is proportional to the electric field amplitude *E* and the phase angle corresponds to the classical phase of the field. The phase-space point represents the idea that, classically, we can measure number and phase simultaneously and with infinite precision. As we have seen above, quantum mechanically this is not so. The Heisenberg Uncertainty Principle (HUP) of Eq. (5) tells us that both phase and intensity cannot be measured simultaneously with infinite precision. For a minimum uncertainty state (MUS), such as a coherent state $|\alpha\rangle$, we have equality in the HUP, as given in Eq. (6a). Then, combined with the Poissonian-statistical distribution of photon number for a coherent state, Eq. (6b), we arrive at the shot-noise limit.

However there are other minimum uncertainty states besides the coherent state. The easiest way to see this is to look again at the representation of the coherent state as a disk in phase space (figure 4). The fact that it is a disk indicates that the fluctuations are the same in all directions, and that the area of the disk is a constant $\mathcal{A}$. The pictogram and the HUP then tells us that any quantum state must have an area greater or equal to $\mathcal{A}$, and that the MUS has an area equal to $\mathcal{A}$. This is, for a coherent state, equivalent to stating the three conditions of Eqs. (6). However, as Caves pointed out, we can relax Eq. (6b) and (6c), while still maintaining the HUP of Eq. (6a). That is, we can decrease $\Delta\varphi$, at the expense of increasing $\Delta n$ at the same time, so that the product $\Delta\varphi\Delta n = 1$ remains constant and the area of the disk remains the same value $\mathcal{A}$. Pictorially this amounts to squeezing the coherent-state disk in the angular direction, while allowing it to expand in the radial direction, as shown in figure 4. The important point is that the area $\mathcal{A}$ of the ellipse remains unchanged so that the HUP is obeyed. However, we decrease phase uncertainty (which we care a great deal about) at the expense of increasing the number uncertainty (which we do not care much about). What is truly amazing is, that it is possible to produce such squeezed states of light in the laboratory, using nonlinear optical devices and ordinary lasers [7]. Without going through the details, which are in a nice review article by Walls and in most quantum optics textbooks [4, 7], let us estimate how well we can do with this squeezing approach.

Again, heuristically, we can ask the question — What is the most uncertainty we can produce in photon number, given that the mean photon number $n$ is a fixed constant, and that we still want to maintain the MUS condition — namely that the area of the ellipse remains a constant $\mathcal{A}$. Intuitively one cannot easily imagine a scenario where the fluctuations in the energy, $\Delta E = \hbar\omega\Delta n$, *exceeds the total energy of the laser beam,* $E = \hbar\omega n$. Hence the best we can hope to achieve is $\Delta E=E$ or, canceling out some constants, $\Delta n=n$. Inserting this expression in the HUP of Eq. (6a), we obtain what is called the Heisenberg limit:

$$\Delta\varphi_{HL} = \frac{1}{n}. \tag{9}$$

Putting back the dimensions we get

$$\Delta\varphi_{HL} \propto \frac{1}{I}. \tag{10}$$

This is exactly the limit one gets with a rigorous derivation using squeezed light in the limit of infinite squeezing. It is called the Heisenberg limit as it saturates the number-phase HUP, and also because it can be proven that this is the best you can do in a passive interferometer with finite average photon number $n$. Converting to minimum detectable displacement we get,

$$\Delta x_{HL} = \frac{\lambda}{n} \propto \frac{\lambda}{I}.$$

What does this mean? Well let us consider our previous example for LIGO where $n=10^{24}$ photons and $\lambda = 1.0$ μm. The quadratic improvement (of infinite squeezing) implies a boost in signal-to-noise by *twelve orders of magnitude!* The minimum detectable displacement is now a flabbergastingly low $\Delta x_{HL} = 10^{-30}$ meters. This is just shy of the Planck length ($10^{-35}$ meters) where classical notions of space and time break down completely and quantum gravity rears its foamy noggin. Talk about your precision measurement!

To understand a bit better why squeezing helps, let us consider the MZI in figure 2. Note that, hitherto, we send light in port A and analyzed what came out ports C and D. What about input port B? Classically there is no light coming in port B, and hence it is irrelevant. Not so quantum electro-dynamically! In his 1981 paper, Caves showed that no matter what state of the photon field you put in port A, so long as you put nothing (quantum vacuum) in port B, you will always recover the SNL. Always! This was a surprise. In 1981 most interferometer experts thought the shot-noise was coming from the statistics of the coherent-state photon field itself — think of the buckshot off the metal plate. What Caves showed was that the SNL arises from the phase noise due to the quantum electrodynamic vacuum fluctuations leaking in the unused port B, *regardless of what you put in port A*. In quantum electrodynamics, even an interferometer mode with no photons in it experiences electric field fluctuations in that mode. While the average intensity is zero for such a vacuum mode, the fluctuations in the average intensity are not zero, and such vacuum fluctuations can be held responsible for a wide array of phenomena such as the Lamb shift, atomic spontaneous emission, and the Casimir force between two conducting mirrors [8].

In the MZI these vacuum fluctuations have another important effect; at the first BS they enter through port B and mix with whatever is coming in port A to give the SNL in overall sensitivity. It becomes clear then, from this result, that the next thing to try would be to plug that unused port B with something besides vacuum. But plug it with what? You can show that if you just put another coherent state $|\beta\rangle$ in port B, you still just get the SNL of Eq. (6c), but now with the total intensity the sum of the two input intensities. If $|\alpha|^2 = |\beta|^2 = n$, you could have gotten the same result by just doubling the power in port A and leaving port B empty again — no real improvement beyond the SNL. Hence, in addition to plugging port B with something, to beat the SNL you must also plug it with something non-classical. It was Caves' idea to plug the unused port B with squeezed light (squeezed vacuum to be exact). That, with coherent laser light in port A as before — and in the limit of infinite squeezing — then the SNL rolls over into the HL and we would be measuring not only the passing of gravity waves but also the graininess of the space-time continuum itself.

Alas, in the laboratory, as might be expected; infinite squeezing is mighty hard to come by. With current technology [9], the expected situation is to sit somewhere between the shot-noise limit (SNL) and the Heisenberg limit (HL) but a heck of a lot closer to the former than the latter. Recent analyses by a Caltech group, on exploiting squeezed light in LIGO, indicates a potential for about a one-order-of-magnitude improvement in a future LIGO upgrade. Not the twelve orders of magnitude that was advertised above, but enough to allow the observatory to sample about eighty times the original volume of Space for gravitational-wave sources. That, for LIGO, is a big deal. In addition, the goal of reaching the true quantum (Heisenberg) limit, which gives us another eleven orders of magnitude to play with, has stimulated other approaches using entangled photon number states, borrowing a trick or two from quantum information theory. We'll look at these digital approaches to quantum optical metrology next.

## 3. Entangled States of Light: The Lowdown on High N00N

I like to say that squeezing is an "analog" approach to quantum optical interferometry, in that the average photon number and the degree of squeezing are continuous variables that can be tuned to any arbitrary non-negative value. There is another approach, exploiting discrete photon number and path-entangled optical states, where the photon number and the degree of squeezing are fixed. This approach, using entangled photon number states, I like to call "digital". The terminology is important since, due to its digital nature, the entangled-number-state approach has recently benefited tremendously from an influx of ideas and experimental techniques originally developed in the context of all-optical "digital" quantum computing [10]. The idea is that an optical quantum computer is a giant optical quantum interferometer with many arms and a large but discrete number of entangled photons flying through it, where the quantum entanglement between photons is exploited to carry out mathematical calculations, which are impossible on any classical computer.

However, our proposed metrological device is also a giant quantum interferometer with entangled photons in it, but here the entanglement is exploited not to solve impossible math problems, but to make ultimately precise measurements not possible with any classical optical machine. The optical quantum computer can be turned into an optical quantum interferometric measuring device, and vice versa. Theoretical and experimental tricks, devised for the former, can be exploited in the latter. Since, for over the past ten years, incredible amounts of effort have gone into the development of quantum computers, we are now able to leverage this research for quantum optical metrology, imaging, and sensing. In fact the metrological applications are much more likely to be realized in the near future. It will take a quantum optical computer with thousands of modes and millions of entangled photons to crack a 1024-bit public crypto key, and hence be of use to the spy agencies. However, a quantum optical interferometer, with only tens of modes, and a few photons, would still be very interesting and important and practical for metrology, imaging, and sensing. The near-term metrological applications will be the bridge technology that will help transition the quantum computer from the few to the millions of quantum bits, or qubits, needed for the computing tasks. Somebody is always willing to pay for a sensor that is an order of magnitude better than the previous generation.

There is a large body of literature on using entangled particles or photons in a Mach-Zehnder interferometer in order to beat the shot-noise limit. The first such proposal was by Bernard Yurke in the context of neutron interferometry [11]. Much of the literature is confusing, in that different researchers exploit different entangled states of light or matter, put these states in the interferometers at different points, and use different detection schemes and signal-to-noise analyses to extract an estimate of the phase resolution. All this confusion makes it difficult to compare the schemes to each other. I have given a short review of what was known prior to the invention of optical quantum computing in my own paper [12] and Viktor Dodonov has provided a more recent overview [13].

In 2001 Emanuel Knill, Raymond LaFlamme, and Gerard Milburn proposed an all-optical, digital, scheme for quantum computing that exploited discrete entangled photon states distributed over the modes of an optical interferometer [14]. This discovery ignited a huge international collaborative research and development program on the optical quantum computer. Not coincidentally, many of those interested in optical quantum computing are also now pursuing quantum optical metrology. It is in this context I would like to discuss recent advances in quantum-entangled photon-state metrology.

In figure 5, we purposely redraw our prototype MZI to look like a quantum computing circuit diagram, in order to better exploit the connection to quantum computing. (For a more detailed elaboration on the connection between quantum computing and quantum interferometry, see our paper on the Quantum Rosetta Stone [15].) What we have done is moved the first BS into the state preparation device, or entangled photon source, on the left, and moved the second BS into the detection scheme on the right. (In the recent literature there is a great deal of confusion on whether the quantum state one is dealing with is introduced before or after the first BS, and I hope to avoid this confusion here with this convention. The point is that, given any two-mode photon state, the input-output BS transform is a simple linear transformation and so the BS does not need to be explicitly discussed as part of the interferometer.)

For the sake of pedagogy, we will at first limit ourselves to a two-mode, path-entangled, photon-number state, a type of Schrödinger Cat state, more commonly called the N00N state. The idea, à la figure 5, is that we have a fixed finite number of photons $N$ that are either all in the upper mode A or all in the lower mode B, but we cannot tell — even in principle — which is which. The state of all up and none down is written $|\text{up}\rangle = |N\rangle_A |0\rangle_B$ and the state of all down and none up is similarly $|\text{down}\rangle = |0\rangle_A |N\rangle_B$. The notation indicates a product state of $N$ photons either in A or B (but not both). There are a number of schemes, now swept all into the entangled-photon source on the left, for producing superpositions of these two states in the form

$$|\text{NOON}\rangle \equiv |\text{up}\rangle + |\text{down}\rangle = |N\rangle_A |0\rangle_B + |0\rangle_A |N\rangle_B, \tag{11}$$

where a normalization constant of $1/\sqrt{2}$ has been dropped for convenience (and convention). This is the high-N00N state, a moniker bestowed by our group due to the happenstance of the notational convenience of choosing the letter $N$ to represent the total photon number. The term "high" indicates that the photon number $N$ is large — at least greater than two. (The competing names of "P00P" and "0NN0" states were summarily discarded at an early stage of the research.) The state is also in the class of Schrödinger-cat states. If the up state is "cat alive" and the down state is "cat dead" a measurement of photon number in either branch A or B will randomly collapse all the $N$ photons into one or the other arms, with a 50-50 probability. As $N$ becomes larger and larger, and the state becomes more and more macroscopic, this becomes spookier and spookier; the whole point of the business with the poor half-dead cat. But all half-dead cats aside — meow! — the N00N state has the interesting property that it is quantum entangled between the two modes and rigorously violates what is known as a Bell inequality for non-classical correlations [16]. This non-locality notion, physically, means that a measurement of photon number $N$ in mode A, also collapses the photon number in mode B to zero, even if detector 𝒜 is on Alpha Centauri and ℬ is on Beta Pictoris. Entanglement is a bookkeeping device that encodes quantum correlations of a non-local nature. Mathematically, entangled states are defined to be quantum states that cannot be written as a product state, like the plain old up or down state given above [17]. The N00N state was first discussed in 1989 by Barry Sanders, who was particularly interested in the Schrödinger-cat aspect and how that affected quantum decoherence [18]. (Big cat states decohere — become dead or alive — more rapidly than small kittens, which just take long naps.) In 2000, my group, then at the NASA Jet Propulsion Laboratory, rediscovered this state in the context of quantum imaging — particularly for quantum lithography [19]. We introduced the term "high-N00N state", which first appeared as a footnote in our group's follow-on paper in quantum metrology [15].

To understand why a N00N state has all the magical properties we will attribute to in the upcoming sections, super sensitivity and super resolution (in particular), we need to understand one more point of quantum interferometry — the difference in behavior between a number state $|N\rangle$ and a coherent state $|\alpha\rangle$ in an MZI. When a coherent state passes through a phase shifter $\varphi$, such as depicted in figure 2, it picks up a phase of $\varphi$. This is a property of a classical monochromatic light beam that coherent states inherit quantum mechanically. However, number states are already highly non-classical states to begin with. Their behavior in the phase-shifter is radically different. When a monochromatic beam of number states passes through a phase shifter, the phase shift is directly proportional to $N$, the number of photons. There is no $n$-dependence in the coherent state, where recall $n$ is the average number of photons. In terms of a unitary evolution of the state, the evolution for any photon state passing through a phase shifter $\varphi$ is governed by

$$\hat{U}(\varphi) \equiv \exp(i\varphi \hat{n}) \tag{12}$$

where $\hat{n}$ is the photon number operator. The phase shift operator can be shown to have the following two different effects on coherent vs. number states [4],

$$\hat{U}_\varphi |\alpha\rangle = e^{i\varphi} |\alpha\rangle, \tag{13a}$$

$$\hat{U}_\varphi |N\rangle = e^{iN\varphi} |N\rangle. \tag{13b}$$

Notice that the phase shift for the coherent state is independent of number, but that there is an $N$ dependence in the exponential for the number state. The number state evolves in phase $N$-times more rapidly than the coherent state. To the right of the phase shifter, in figure 5, the N00N state evolves into,

$$|N\rangle_A |0\rangle_B + |0\rangle_A |N\rangle_B \to |N\rangle_A |0\rangle_B + e^{iN\varphi} |0\rangle_A |N\rangle_B, \tag{14}$$

which is the origin of the quantum improvement we are seeking. If one now carries out a simple measurement scheme in the $N$-photon detecting analyzer we have,

$$M_{\text{NOON}}(\varphi) = I_A \cos(N\varphi), \tag{15}$$

which is the green curve in figure 6. The N00N signal (green) oscillates $N$ times as fast as the coherent state (red). Two things are immediately clear. The distance between peaks goes from $\lambda \to \lambda/N$, which is the quantum lithography effect — we now beat the Rayleigh diffraction limit of λ by a factor of $N$. This sub-Rayleigh-diffraction-limit effect is now commonly called "super-resolution". Even better, the slope of the curve at the horizontal axis crossing point gets larger, also by a factor of $N$, and our minimal detectable phase, given by Eq. (4), goes down. However the signal $M$ for this N00N state is not the same as for the coherent state scheme, as we are now counting photons $N$ at a time. And it turns out then that $\Delta M_{\text{NOON}} = 1$ for the new scheme, and then Eq. (4) gives:

$$\Delta\varphi_{\text{NOON}} = 1/N, \tag{16}$$

which is precisely the Heisenberg limit of Eq. (9). This Heisenberg limit, or the beating of the shot-noise limit, is now commonly called "super-sensitivity". To see why N00N states have this property, without a detailed analysis of the detection scheme [20], let us make the same heuristic argument as before. For a minimum uncertainty state, we expect the best performance when the uncertainty in photon number is largest, but here it is the uncertainty in number distributed over the two-photon paths, A and B. In N00N states we are completely uncertain if all $N$ photons are in mode A (and none in B) or if all the $N$ photons are in mode B (and none in A), where A and B may be light years apart. It does not get any more uncertain than that!

Hence, as before, we conclude $\Delta N = N$, and plugging into the Heisenberg relation, Eq. (6a), we immediately arrive again at the Heisenberg limit, Eq. (9). This is the same argument we made for squeezed states, in the limit of infinite squeezing, but infinite squeezing is not required to demonstrate this effect digitally with N00N states, at least up to about $N = 4$. We will discuss some of the recent experiments in an upcoming section.

**4. How to Make a High-N00N State?**

While there are other states besides N00N states that exhibit super resolution (beating the Rayleigh diffraction limit) and other states besides N00N states that exhibit super sensitivity (beating the shot-noise limit), the N00N state accomplishes both in a single, easy to understand, quantum mechanical package. For this reason much of the recent theoretical and experimental work has revolved around these states. For $N = 1$ and $N = 2$ (low-N00N) it is fairly easy to make such states with non-classical sources of photon number states of either the form $|1\rangle_A|0\rangle_B$ or $|1\rangle_A|1\rangle_B$, that is one photon in mode A and none in B, or one photon in each of mode A and B. The standard approach utilizes spontaneous parametric down-conversion (SPDC), where an ultraviolet (UV) photon is down converted into a pair of number states [13]. The effect of a simple beam splitter transformation on these states [4], is to convert them to low-N00N states, as follows,

$$|1\rangle_A|0\rangle_B \xrightarrow{BS} |1\rangle_{A'}|0\rangle_{B'} + |0\rangle_{A'}|1\rangle_{B'}, \tag{17a}$$

$$|1\rangle_A|1\rangle_B \xrightarrow{BS} |2\rangle_{A'}|0\rangle_{B'} + |0\rangle_{A'}|2\rangle_{B'}, \tag{17b}$$

where Eq. (17a) shows that a single photon cannot be split in two, and Eq. (17b) is illustrative of the more subtle Hong-Ou-Mandel effect — if two single photons are incident on a 50-50 beam splitter they will "stick" and both photons will go one way or both will go the other way, but you never get one photon out each port [4]. The way to understand this is that the probability of the transition $|1\rangle_A|1\rangle_B \xrightarrow{BS} |1\rangle_{A'}|1\rangle_{B'}$ completely cancels out due to destructive interference, and the transition indicated by Eq. (17b) adds up, due to constructive interference. So it is relatively easy, once you have a source of single photons, to create low-N00N states. The challenge is then, how to go to high-N00N?

In the SPDC process, particularly if the UV pump laser is very intense, the output produces, with high probability, un-entangled number states of the form $|N/2\rangle_A|N/2\rangle_B$, where we take $N$ to be even for this example. These states are called twin number states and are the basis for a bunch of super-sensitivity schemes. From the twin-number state, one can produce number states of the form $|N/2\rangle_A|0\rangle_B$ by checking to see if N/2 is in mode B and then allowing what is in mode A into the interferometer, in a process called heralded number-state production. (None of these things are particularly easy to implement in the lab, but at least it is a start.) So, in a generalization of Eq. (17b), we might guess the transformation,

$$|N/2\rangle_A|N/2\rangle_B \xrightarrow{BS?} |N\rangle_{A'}|0\rangle_{B'} + |0\rangle_{A'}|N\rangle_{B'}, \tag{18}$$

and hence have an easy road to high-N00N. But, alas, this idea is too good to be true. In what is now called the generalized Hong-Ou-Mandel effect, the actual 50-50 beam splitter performs,

$$|N/2\rangle_A |N/2\rangle_B \xrightarrow{BS}$$
$$c_N |N\rangle_{A'} |0\rangle_{B'} + c_{N-2} |N-2\rangle_{A'} |2\rangle_{B'} + \ldots + c_{N-2} |2\rangle_{A'} |N-2\rangle_{B'} + c_N |0\rangle_{A'} |N\rangle_{B'}, \quad (19)$$

where the constants $c_n$ are probability amplitude weight factors. We can see that at each end is the desired N00N-state component, but there is much POOP in the middle that is undesirable. Actually, the N00N components at the two ends, remarkably, have the highest probability of showing up. This state, complete with POOP, can be exploited to exhibit super-sensitivity, but only the pure N00N state can actually hit the Heisenberg limit [20]. So to extract the pure N00N state at the two ends of this distribution, we need to somehow get rid of the POOP in the middle. One approach is a quantum pooper-scooper; something that removes the POOP after the fact [21]. Another approach is make sure the POOP is not there to begin with, but this requires something besides an ordinary 50-50 beam splitter — something we call a "magic" beam splitter [15]. Much of the discussion that follows is devoted to magic BS.

One of the first proposals for making a magic BS was introduced in 2002 by Gerry and Campos (GC), motivated by the application of high-N00N states to lithography and metrology [28]. The GC scheme is shown in figure 7. The idea is to make a kind of quantum computing gate (a Fredkin gate) so that a single photon in the upper MZI-1 controls the phase shift in the lower MZI-2. A nonlinear material called a cross-Kerr phase shifter couples the two MZIs. For very, very, very good Kerr materials, the presence of even a single photon in the upper MZI-1 can cause a phase shift of $\pi$ in the lower MZI-2. Let us consider first the lower MZI-1 and assume the box labeled "Kerr" is an ordinary phase shifter. If the phase shift is set to zero then one can show that if $N$ photons are incident from the left in mode A and none in mode B then all $N$ will exit out port D. This is just the generalization of the balanced MZI effect we discussed above, where D is still the bright port and C the dark port, giving output state $|0\rangle_C |N\rangle_D$. Exactly as before, if we put a $\pi$ phase shifter in one arm then, all $N$ photons exit out the upper port C and none out port D, giving output state $|N\rangle_C |0\rangle_D$

OK, now here is where it gets weird. If I could somehow engineer a phase shifter that could simultaneously be in a superposition of zero and $\pi$ phase shifts, I would get a superposition of $|N\rangle_C |0\rangle_D$ and $|0\rangle_C |N\rangle_D$, but that is the N00N state $|N\rangle_C |0\rangle_D + |0\rangle_C |N\rangle_D$. The idea of Gerry and Campos is to add the Kerr material and the upper MZI-1 to make this whacky, Jekyll and Hyde, phase shifter. The (quantum) mechanics is as follows. In the upper MZI-1 we send in one photon in port a. From Eq. (17a), we see that the single photon is either in the upper path or the lower path in the MZI-1, after the first beam splitter. If it is in the lower path of MZI-1, the interaction with Kerr gives a $\pi$ phase shift for the lower machine, MZI-2. If the single photon is in the upper path of MZI-1, then there is no induced phase shift in the lower machine, MZI-2. However, since the single photon in the upper MZI-1 is in an equal superposition of upper and lower paths, it induces the requisite superposition of zero and $\pi$ phase shift in the MZI-2 — and we have our N00N state. The two, Kerr-coupled, MZI interferometers act as a single magic beam splitter!

However, Houston, we have a problem. The nonlinear Kerr material that you buy at your local Kerr dealership has a very, very, very, very, …, very small response at the single photon level [22]. If $N = 10$ this scheme will produce a superposition of a zero and a $10^{-20}\pi$ phase shift; that is it will come up 20 orders of magnitude too short for making a good N00N generator. There are two well-known ways to boost the Kerr effect: put in an optical micro-cavity around the atoms in the Kerr material, or coherently lock the atoms together in an approach known as electromagnetically induced transparency (EIT). Both roads have their complication and technical challenges. There is, however, a third path.

Here is where the N00N-atic fringe takes some hints from the world of optical quantum computing. I have mentioned that the Gerry-Campos idea is based on an optical Fredkin gate, a sort of a single photon transistor. If such a device could be made easily, it would be a quick and easy road to the all-optical quantum computer. In fact, older optical approaches to quantum computing typically involved boosted Kerr nonlinearities with cavities [23] or with EIT schemes [24]. Enter, *deus ex machina,* in 2001, the all-linear optical approach to quantum computing [14]. In this approach, the Kerr nonlinearity is replaced with additional, ancillary, mirrors, phase shifters, beam splitters and — most importantly — detectors. The idea is that the detection process in the ancillary devices induces an effective Kerr nonlinearity [25]. While still not perfect, the effective Kerr produced this way can be much stronger than the off-the-shelf block of Kerr material. So instead of working one time in $10^{20}$, our new device works one time in 10, which is a 19-order-of-magnitude improvement. Our group, then at the NASA Jet Propulsion Laboratory, proposed the first high-N00N generation scheme based on all linear-optical devices [26]. The scheme is shown in figure 8. The idea is to make the upper and lower mirrors into additional beam splitters and to put detectors just past them. We launch into the system from the left the twin-number state $|3\rangle|3\rangle$, which can be generated from highly pumped SPDC. (We drop the subscripts A and B with the convention that the first state is in the upper path and the second state is in the lower path.) After the first 50-50 BS we have, from the generalized Hong-Ou-Mandel effect, the coherent superposition of the states $|6\rangle|0\rangle$, $|4\rangle|2\rangle$, $|2\rangle|4\rangle$, and $|0\rangle|6\rangle$. Now the trick! We monitor the upper and lower detectors and check for the case when both detect one (and only one) photon in each detector. This is a heralded process, we will only consider further processing on events where this joint detection occurs. If both detectors click "one" then we know a total of two photons are missing from the interferometer, one from the upper arm and one from the lower arm. Hence, the contributions $|6\rangle|0\rangle$ and $|0\rangle|6\rangle$ collapse out (vanish) from the state. The zero can never give a click. Yet we still know two photons are missing, one from each arm, so after the intermediate detection we have only the contributions $|3\rangle|1\rangle$ and $|1\rangle|3\rangle$ left. It is easy to show, in a reverse Hong-Ou-Mandel effect, that when these are run through the final BS, out comes the $|4\rangle|0\rangle + |0\rangle|4\rangle$ N00N state. Now if one optimizes the reflectivity of the beam splitters and everything is perfect, then the N00N state only emerges about one time in 10. However in the Gerry-Campos scheme a N00N state emerged only one time in $10^{20}$. Hence, our linear-optical scheme represents a 19-order-of-magnitude improvement over the original GC scheme and — even better — can be executed in the lab with fairly simple equipment.

Our N00N-generating scheme can be concatenated — or stacked — to produce N00N states of arbitrarily high $N$ [27]. However, at least initially, all such schemes produced N00N states with a probability that scaled exponentially badly as $N$ increased. The larger and larger you made $N$, the far, far, far more likely it was that you would get only POOP. Recently there has been a great deal of work on the development of schemes which do much better — in quantum computing lingo — schemes that produce N00N "efficiently" [21, 29, 30]. So we are well along the digital quantum-computer-paved road to super resolution and super sensitivity — at least in theory. But what about the experiments? We shall talk about these next.

## 5. Showdown at High N00N: The Experiments

Since the phase shift imparted in the single-photon N00N state of Eq. (17a) is identical to that of a coherent state, the first interesting case comes about when $N = 2$. As discussed above, this can be generated via the Hong-Ou-Mandel effect, indicated in Eq. (17b). The standard setup is to send a bright beam of UV photons onto a $\chi^{(2)}$ nonlinear crystal, which is typically a birefringent crystal that converts a UV photon into two red daughter photons. By playing tricks with the polarization of the UV beam and the orientation of the crystal, you can get the daughter photons to emerge in two separate paths (or modes) as shown in figure 9. The output is effectively the $|1\rangle_A |1\rangle_B$ state, which is then transformed by the BS, as per Eq. (17b), into the low-N00N state with $N = 2$. A series of experiments illustrating this low-N00N production appeared around 1990 [31, 32, 33].

In 1998, Kuzmich and Mandel provided an experimental demonstration showing that such states beat the shot-noise limit (super sensitivity) [34]. The explicit demonstration of the doubling of the effective wavelength (super resolution) was made in 1999 [35], and a mockup demonstration of the super-resolving application to sub-Rayleigh lithography was carried out in the group of Yanhua Shih in 2001 [36]. Then in 2004 the $N = 2$ barrier was breached in two back-to-back papers appearing in *Nature;* the group of Steinberg demonstrated super resolution for $N = 3$, and the group of Zeilinger did so for $N = 4$ [37, 38, 39]. These 2004 experiments exploited our idea of inducing *effective* Kerr nonlinearities extracted from the photodetection process, as shown in figure 8. In fact, the Zeilinger N00N experiment, and the Zeilinger linear optical quantum computing experiment [40], were performed with the same apparatus. This feat shows the close connection between the fields of quantum computing and quantum metrology, even in the experimental realm.

While these 2004 experiments were very suggestive, it was Steinberg's group who pointed out very clearly that super-resolution does not imply super-sensitivity [38]. Just because the graph wiggles $N$ times as fast does not mean you can beat the shot-noise limit. The issue is that, in any experiment, there are imperfect photodetectors and other sources of loss, and the visibility of the detection curves in figure 6 is less than one. Here the visibility is defined as,

$$V = \frac{M_{max} - M_{min}}{M_{max} + M_{min}}. \tag{20}$$

For a lossless system, which is what is calculated for figure 6, $M_{max} = I_A$ and $M_{min} = -I_A$ and so $V = 1$. However, as we shall see in section 7, below, N00N states are typically exponentially more susceptible to loss than are coherent states. In figure 10 we indicate the same experiment as in figure 6, except now with 3 dB (50%) loss in on arm of the interferometer. One can see that the green N00N curve ($N = 3$) still exhibits the high-frequency wiggles (super resolution) but now the maximum slope of the green curve at the $\pi/2$ crossing point is only as steep as that of the coherent state red curve, and we do about as bad — or worse! — than the shot-noise limit, as per Eq. (6c). The visibility for the coherent state (red curve) is 50% but that of the $N = 3$ N00N state (green curve) is $(0.5)^3 = 0.125$ or about 13%. Hence it is possible to have super resolution — without super sensitivity — if the system has sufficiently high photon losses or, equivalently, if the visibility is sufficiently small.

In 2007, a number of very interesting experiments occurred in quantum optical metrology. First the group of Andrew White in Australia demonstrated $N = 6$ super resolution in a process which used classical photons with a nonlinear $N$-photon detection scheme. While interesting, such an approach can never achieve super-sensitivity due to its semi-classical nature, as the authors themselves pointed out [41]. The first demonstration of both super sensitivity and super resolution in a single experiment also appeared in late 2007 in a collaborative Japanese and UK experiment [42]. This was the first experiment to beat the shot-noise limit, using N00N states, with $N > 2$. However the results still were not quite *at* the mythical Heisenberg limit. (You can be somewhere between the shot-noise limit and the Heisenberg limit, and again this is a visibility issue related to photon loss and detector inefficiency.)

At this point something came out of the sky like a bolt from the clear-blue sky at high noon. Another Australian collaboration, of the groups of Geoffrey Pryde and Howard Wiseman, produced an experiment that, although not using N00N states *per se,* produced $N$ entangled *single*-photon states of the form of Eq. (17a), and recycled them through the MZI in a feedback-loop implementation of a quantum computing protocol known as the Kitaev phase estimation algorithm [43, 44, 45]. The effective $N$ in this experiment, in N00N-state currency, was a mind-boggling $N = 378$. (My favorite number!) The idea is that instead of making such a large N00N state (cat), they make 378 passes through the interferometer, with feedback, using the low $N = 1$ N00N (kitten) state. Since N00N states are equally entangled and violate a Bell inequality for any non-zero value of $N$, the trade off is that of a complicated N00N-state generating scheme with a less complicated single photon detection scheme with some electronic feedback. Such a protocol is, surprisingly, easier to implement than the high-N00N approach, and arrives at the same performance in sensitivity scaling, the Heisenberg limit. This Australian experiment, once again, illustrates the close connection between quantum optical computing and quantum metrology, as it achieves super sensitivity by running a quantum computational algorithm. A quantum computer is nothing more than a quantum sensor, and hence one can design a quantum sensor by exploiting ideas from quantum computing.

However, there is one remaining bone here I would like to pick: This Australian paper claims that their single-photon experiment hits the Heisenberg limit without requiring entanglement [45]. That statement simply is not true! The circulating single photons in the interferometer have the form of Eq. (17a), and these are perfectly respectable entangled states, which can even be shown to maximally violate a Bell inequality for non-locality [Wildfeuer2007]. Perhaps the authors meant to say that the single photons emitted from the source — *before the beam splitter* — are not entangled; but single photons are highly non-classical and become entangled upon passing through an ordinary beam splitter, as discussed above. (In the experiment, the single photons are generated from spontaneous parametric down conversion — a notorious source of entangled photons!) The role of quantum entanglement and non-locality in experiments exhibiting super sensitivity is not exactly clear, at least to me. However, all experiments to date that display super sensitivity appear to require an entangled resource, and I conjecture that entanglement is always a requirement for super sensitivity. Super resolution is another matter. The paper by Resch, et al. [41], explicitly shows super resolution, likely cannot show super sensitivity, and I will not try to quibble and claim that there is entanglement hidden in this experiment someplace. It would, however, be very interesting to clarify this issue: Does super sensitivity always require entanglement and quantum non-locality?

## 6. Quantum Imaging and Sensing

Quantum imaging is a new sub-field of quantum optics that exploits quantum correlations, such as quantum entanglement of the electromagnetic field, in order to image objects with a resolution (or other imaging criteria) that is beyond what is possible in classical optics. Examples of quantum imaging are quantum ghost imaging, quantum lithography, and sub-Rayleigh imaging [46]. Of particular interest for this article are lithography and sub-Rayleigh imaging and sensing. In our 2000 paper on quantum lithography [19], our group pointed out that N00N states had the capability to beat the Rayleigh diffraction limit by a factor of *N*. This super resolution feature is due to the high-frequency oscillations of the N00N state in the interferometer, as illustrated in figure 6.

For the quantum lithography application, the idea is to realize that if one has an *N*-photon absorbing material, used as a lithographic resist, then these high-frequency oscillations are written onto the material in real space and are not just a trace on an oscilloscope. Mathematically, the *N*-photon absorption and the *N*-photon detection process have a similar structure, that is,

$$\delta(\varphi) = \langle \mathrm{NOON} | (\hat{a}^\dagger)^N (\hat{a})^N | \mathrm{NOON} \rangle = 1 + \cos(N\varphi), \tag{20}$$

which is just the expression of Eq. (15), with the intensity scaled out. From figure 6, we see in the green curve this oscillates *N* times faster as if we were using single photons, or coherent light, as in the red curve. Recall that, for our MZI, we have $\varphi = kx = 2\pi x / \lambda$, where x is the displacement between the two arms. For lithography $x$ is also the distance measured on the photographic plate or lithographic resist. If we compare the classical resolution to the N00N resolution we may write, $N\varphi_{NOON} = \varphi_{classical}$, which we can solve for,

$$\lambda_{NOON} = \frac{\lambda_{classical}}{N}. \tag{21}$$

Written this way, we can say the effective wavelength of the $N$ photons bundled together $N$ at a time into the N00N state is $N$ times smaller than the classical wavelength. This is another way to understand the super-resolution effect. The $N$ entangled photons conspire to behave as a single classical photon of a wavelength smaller by a factor of $N$. Since the Rayleigh diffraction limit for lithography is couched in terms of the minimal resolvable distance $\Delta x = \lambda_{classical}$, then we have $\Delta x_{N00N} = \lambda_{N00N} = \lambda_{classsical}/N$.

So let us take an example. Suppose I want to do lithography with red photons of wavelength 500 nm. Then I am limited to image objects on the resist to a separation of 500 nm. That means wires and transistors can be no closer than 500 nm on the resultant computer chip I make. But if I want to make my computer chips faster with more memory, I need to put more features on the chip. I can only do this if I can make the features smaller and pack them closer together. So it is advantageous to go to a 50 nm separation, which classically would require light of wavelength 50 nm, or x-rays. If I reduce the feature size and spacing by a factor of 10 then I can put ten times more gizmos on the chip in the *x* direction and another ten times more objects in the *y* direction and produce a chip with a 100 times more transistors on it. This is in fact what the semiconductor industry proposes, using light at ever shorter wavelengths to make features of ever smaller sizes with ever smaller separation.

Current commercial lithography exploits extreme ultraviolet light of around 100 nm and plans are to go to x-ray in the future. The problem is that the lithography system for x-ray cannot use the same lenses, mirrors, and other imaging devices as did the optical system and so each reduction in wavelength involves a huge cost in technology and hardware investment. But what if I could etch 50 nm sized features using 500 nm wavelength photons by exploiting quantum entanglement? This is the promise of quantum lithography. The idea is that we can keep all the simple, inexpensive, optical imaging systems that work fine at 500 nm and wave our quantum magic wand, entangling the photons 10 at a time into *N = 10* N00N states, and produce quantum states that behave as if they were photons of a 50 nm wavelength, without all the extra trouble and cost. We have the equivalent of a tabletop x-ray laser! The holdup with this quantum approach, which may be insurmountable, is that it has been very difficult to make *N*-photon absorbing resists that are efficient and practical for quantum lithography. So, currently, quantum lithography appears to be road kill on the semiconductor road map, but there is still hope that we are not dead yet and that some smart chemist will produce lithographic resists optimized for the *N*-photon absorption application. However, no real demonstration of quantum lithography has been had, so far, due to the resist problem [47].

However, the pessimism of quantum lithography aside, there is more to quantum imaging than just lithography. Current experiments, mentioned above, avoid the whole issue of the need for *N*-photon absorbing materials by exploiting *N*-photon coincidence detection. Such coincidence detection also gives rise to Eq. (21), but instead of some magical designer molecule tailored to absorb *N* photons, we now just need *N*, good, photocounters. These are much easier to come by. A particular application of this more general idea of quantum imaging has been seen in quantum coherence tomography [48]. In this experiment, they image a phase object sample placed in one arm of the interferometer, using entangled photons in an *N = 2* N00N state. They see not only the factor of two improvement in resolving power, predicted by Eq. (21), but also, as a bonus, they get a dispersion cancellation in the imaging system due to frequency entanglement between the photons. Scaling up to higher *N* would require brighter sources of N00N states, and more sensitive arrays of photo-detectors, but no pesky *N*-photon absorbing substrate.

Current experiments on N00N states have use rather dim sources of entangled photons, from UV pumped $\chi_{(2)}$ crystals in spontaneous parametric down conversion set ups, as indicated in figure 9. For bright sources of N00N states, one can turn to optical parametric amplifiers (OPA), which is the same setup as figure 9, but in which we crank up the pump power. In this regime of high gain, the creation of entangled photon pairs of the form of Eq. (19) occurs, but we have many, many, pairs and the output can be written,

$$|\text{OPA}\rangle = \sum_{n=0}^{\infty} a_n |n\rangle_A |n\rangle_B, \tag{22}$$

where the probability of a large twin-number state $|N\rangle_A |N\rangle_B$ is given by $|a_N|^2$, which can be quite large in the limit of high pump powers. Passing the OPA state through a 50-50 beam splitter, gives the generalized Hong-Ou-Mandel effect, term by term, so that we get,

$$|\text{OPA}\rangle \xrightarrow{BS} \sum_{n=0}^{\infty} \sum_{m=0}^{n} c_{nm} |2n-2m\rangle |2m\rangle, \tag{23}$$

where again the coefficients $c_{nm}$ can be quite large for high pump powers. Taking the term *n = 1* we immediately get the *N = 2* N00N state from the regular Hong-Ou-Mandel effect. For larger *n ≥1*, we find that there is always a large N00N component squirreled away among the POOP. A concern might be that if we pump the heck out of the sucker (increase the gain) then the POOP might overwhelm the N00N and hence that that as we increased the output flux (good for lithography) the N00N oscillation visibility would disappear (bad for lithography). Somewhat miraculously, this is not the case. For an *N = 2* absorber, the visibility of the *N = 2* N00N oscillations was predicted to saturate at a visibility of 20% [50, 51], as opposed to the 0% that might have been expected by the naive argument. This 20% visibility is more than enough to exploit for lithography and imaging, and has recently been measured in a recent experiment in the group of DeMartini [52], in collaboration with our activity at Louisiana State University (LSU). The moral of this story is that bright sources for super-resolution imaging are available and have performances that can be exploited in a practical set up.

## 7. Quantum Remote Sensing and Photon Loss

Improvements in optical metrology and imaging have a natural application in the realm of optical remote sensing. Our activity at LSU was the first to suggest that, by exploiting entangled photon states, one could engineer resolution and sensitivity breakthroughs over classical optical sensors, such as in coherent optical laser interferometric radar (LIDAR) [53], or in sensor miniaturization [54]. Of course, when one thinks of N00N states propagating over distances of kilometers through, say, the atmosphere, then photon scattering and loss and other issues, such as atmospheric turbulence become an issue that are not apparent in a table top quantum interferometry demonstration, where the environment is very well controlled. In the past year a number of groups, including our own, has been investigating the effect of photon loss or absorption on the super-resolving and super-sensitivity of N00N state interferometry.

As discussed in Sec. (5), above, the primary issue associated with photon loss is that that the visibility of the interference pattern decreases, and that of the N00N state pattern decreases more rapidly than that of the single photon or coherent state interferometer. Hence, when the loss is sufficiently high, the slope of the N00N oscillations in figure 6 decreases to the point that, as far as super-sensitivity is concerned, we do worse with N00N states than with either single photons or coherent states [55]. Let us see why this is so.

Consider Eqs. (13), describing how coherent states and number states behave upon passing through a phase shifter. It is typical to model loss in the phase shifter by making the substitution $\varphi \rightarrow \varphi + i\gamma$, where $\gamma$ is the rate at which photons are absorbed, say, by impurities in the glass. For the sake of this discussion, we assume that all of the loss in the system, including detector inefficiency, is concentrated in this single parameter. Immediately we see that the effect of this loss in Eq. (13) is to produce an exponential loss factor that depends on $N$, for number states,

$$|\alpha\rangle \xrightarrow{\varphi+i\gamma} e^{-\gamma} e^{i\varphi} |\alpha\rangle, \tag{24a}$$

$$|N\rangle \xrightarrow{\varphi+i\gamma} e^{-N\gamma} e^{iN\varphi} |\alpha\rangle. \tag{24b}$$

Typically we have $\gamma = gL$, where $g$ is the loss per unit length and $L$ the distance traveled through the lossy medium. The exponential dependence of the loss in the coherent (classical) state of Eq. (25a) is called Beer's Law for optical absorption. We see that for $N$-photon number states, Eq. (25b), we have a super-exponential behavior, or what we call super-Beer's Law. Guinness stout notwithstanding, super Beer is bad news for N00N states, as it implies that they are much more fragile in a lossy environment than a classical coherent state. The effect of super Beer's law is plotted in figure 10, where we have chosen $\gamma = \ln 2$, which corresponds to 3dB or 50% loss in one arm of the interferometer. We can see that visibility for the coherent state, plotted in red, decreases by a factor of 2. However the visibility of the $N = 3$ N00N state, plotted in green, decreases by a factor of $2^3=8$. This is close to the breakeven point where the sensitivity of the N00N-state measurement rolls back to the old shot-noise limit, which can be understood in that the maximum slope of the green and red curves, at the point $\varphi = \pi/2$, is about the same. As per the minimal detectable phase estimator of Eq. (4), this implies the coherent and N00N state are about equal in sensitivity for this amount of loss. Any more loss and the N00N state would actually do worse than the coherent state!

What can we conclude from this analysis? When the N00N state was first introduced in 2000, it became an icon for quantum optical metrology. Here was a single, entangled, quantum state that, in a fairly intuitive way, could be seen to give rise to super resolution and super sensitivity all in one package. In the year 2000, it was remarkable that there was any photon state that could do these things, and nobody was thinking about using N00N states in the context of remote sensing at that time. Times have changed. The U. S. Defense Advanced Research Projects Agency (DARPA) now has a program that explicitly seeks to answer the question: Can quantum states of light be used to get super resolution and super sensitivity in a remote sensing environment where losses can be severe? Hence understanding the quantum mechanical underpinning of the super-Beer's law becomes important, as well as addressing ways to mitigate the loss but maintaining the feature of *super-quantum-phase realistically extracted à la photons.* In spite of the loss having such an adverse effect on N00N states, we can imagine some applications where the distances are not too large, and the losses not too big, so that N00N states could give a definite signal-to-noise advantage. Also, although there are proofs that N00N states are optimal for phase resolution in the absence of loss [20], these proofs leave open the possibility that there are other states of light with super-resolving or super-sensing capability that are more robust, or that are some sense, immune to loss.

## 8. Conclusions

There is much more to quantum optical metrology than N00N states. So I may be faulted for not covering enough. I felt, however, that I could not readily cover everything in this review, so I focused on these N00N states in part because I am most familiar with them, and also in part because I think they provide a simple and intuitive way to see how entanglement in quantum optics can be exploited in metrology, with consequent applications to imaging and sensing.

However, for completeness, I thought I should mention some other things that have been going on in the field of quantum optical metrology. For example, Alan Migdall at the US National Institute of Technology has proposed and implemented a quantum optical technique for calibrating the efficiency of photo-detectors using the temporal correlations of entangled photon pairs [56]. This, to my mind, was one of the first practical applications of quantum optics to optical metrology, and has produced a technique to calibrate detectors without the need for an absolute standard. In quantum imaging, there is more going on than just super resolution with N00N states. There are some nice reviews on the field of quantum imaging [46, 57], but I can touch on the highlights. There has been a tremendous amount of work in recent years on so-called "ghost imaging" [58]. This effect exploits the temporal and spatial correlations of photon pairs, also from spontaneous parametric down conversion, to image an object in one branch of the interferometer by looking at correlations in the coincidence counts of the photons. The cute thing here is that there is no image in the single-photon counts in either arm, but only in the double photon counts in both arms. The image is in a sense stored nonlocally. There is the quantum-coherence-optical-tomography-microscope experiment at Boston University, which exploits the super-resolution effect we have discussed, but also utilizes the entanglement of the photon pairs in frequency to mitigate dispersion effects that typically limit the resolution of classical optical tomography [48]. This dispersion cancellation business goes back to the original work of James Franson [59] and Aephraim Steinberg [60], from the early 1990's. Quantum-entangled dispersion cancellation has been harnessed, at least in theory, by the group of Seth Lloyd at MIT, who proposed a quantum optical clock synchronization protocol that eliminates the timing jitter of optical pulses that are transmitted through a fluctuating atmosphere [61]. These atmospheric fluctuations, similar to those that cause the twinkling of the stars, are currently the limiting source of noise in the Global Positioning System.

My own philosophical take on quantum optical metrology is to say that there is no such thing as classical optics. All optical sensing and imaging systems are quantum mechanical, in that photons must be invoked and the quantum origins of signal to noise considered, at least at some level of accuracy. The question then is, at this level, what is the best you can do according to the laws of quantum mechanics? Re-looking at the whole field of optical metrology, from a quantum point of view, opens up all sorts of possibilities from the quantum bag of tricks that could improve optical sensors. We are just beginning to understanding what role the quantum features play, and how such features could be exploited, in theory and in practice. Boy, has the fun just begun, or what!?

**Acknowledgements**

Many people have contributed to the work that I cite here, but I particularly would like to acknowledge Pieter Kok and Hwang Lee in our quest to understand high-N00N states and their implications for quantum optical metrology. The author would also like to acknowledge support from the U. S. Army Research Office, the U. S. Intelligence Advanced Research Projects Activity, and the U. S. Defense Advanced Research Projects Agency.

*Jonathan P. Dowling* has been the Co-Director of the Hearne Institute for Theoretical Physics since 2006 and Hearne Professor of Theoretical Physics in the Department of Physics and Astronomy since 2004, both at Louisiana State University. Previously he was a Principal Scientist at the NASA Jet Propulsion Laboratory, California Institute of Technology, and a Research Physicist at U. S. Army Aviation and Missile Command. His interests include quantum optics, foundations of quantum mechanics, and quantum technology. Dr. Dowling has over 125 journal publications and is a co-editor on a number of special journal volumes in quantum technology such as *Quantum Imaging* 2006 (J. Mod. Opt.), *Single-Photon: Detectors, Applications, and Measurement Methods* 2004 (J. Mod. Opt.), *Quantum Dots For Quantum Computing* 2002 (Superlattice Microstructures). He is also editor of *Electron Theory and Quantum Electrodynamics — 100 Years Later* 1997 (Plenum Press). Dr. Dowling is a Fellow of the Optical Society of America and member of the American Physical Society. He is a recipient of the Willis E. Lamb Medal for Laser Science and Quantum Optics (2002), the NASA Space Act Award (2002), and the U. S. Army Research & Development Achievement Award (1996).

**Figure Captions**

All figures created in MS PowerPoint 2004 for Macintosh

Figure 1. In a Michelson interferometer, laser light is incident on a 50-50 beam splitter and the two resultant beams are launched into two perpendicular arms, which in LIGO have a length of 4 km. On their round trips between the beamsplitter and the mirrors, the laser beams accumulate a phase difference $\varphi$, which is proportional to the displacement between the two arms caused by a passing gravity wave. The beams then recombine at the beamsplitter and emerge from the lower port to be detected. The detection port is typically balanced to be "dark" so that any light that is emergent here indicates a nonzero arm displacement.

Figure 2. A Mach-Zehnder interferometer is an "unfolded" Michelson interferometer where two separated beam splitters play the role of the single one. Laser light in port A is split by the first 50-50 beam splitter, reflected off two mirrors, and then accumulates a phase difference, that again is proportional to the path difference between the upper and lower arms. The light recombines at the second beam splitter and emerges in ports C and D. Typically, for a balanced interferometer, port D is the dark port. Hence any light emergent here is indicative of an arm displacement and can be detected by the two detectors and the analyzer.

Figure 3. In a typical classical Mach-Zehnder analyzer, we subtract the optical intensities at the detectors $\mathcal{C}$ and $\mathcal{D}$. The resultant plot is shown here as a function of the phase shift $\varphi = kx$, where $x$ is the arm displacement to be measured. The difference intensity $M$ is periodic with period $2\pi$. The minimal detectable displacement, $\Delta x$, is limited by the fluctuations in the optical intensity, $\Delta M$. These fluctuations are quantum mechanical in nature.

Figure 4. In this phase-space diagram, the major and minor axes of the ellipses depict fluctuations of various states. Fluctuations in the radial direction correspond to intensity and those in the angular direction phase. A classical state is a point and has no fluctuations. A coherent state, that of a typical laser, is a disk and has fluctuations equal in intensity and phase. Also shown is a phase squeezed state, which has fluctuations decreased in the angular (phase) direction, at the expense of increase fluctuation in the radial (intensity) direction. Such a phase-squeezed state can be used to beat the shot-noise limit.

Figure 5. Here is a schematic depiction of a two-mode optical interferometer as a source, phase shifter, and detector.

Figure 6. Here is a comparison of the detection signal of a coherent state (red) and a N00N state (green). The N00N state signal oscillates $N$ times as fast as the coherent state (super resolution) and has maximum slope that is $N$ times as steep (super sensitivity). (Here we choose $N = 3$.) The effect is as if the N00N state was composed of photons with an effective wavelength of $\lambda/N$ instead of $\lambda$.

Figure 7. Here we show a "magic" beam splitter or a N00N state generator. In the lower interferometer if *N* photons enter upper port A they will always emerge in lower port D for a balanced Mach-Zehnder. However, if we couple the lower interferometer to the upper one, via a strong cross Kerr nonlinearity, a single photon in the lower branch of the upper interferometer causes a p phase shift, directing all N photons to emerge out port C. If the upper device is also an interferometer, one can arrange a superposition of zero and one photons in the Kerr box, giving rise to a superposition of a 0 and $\pi$ phase shift. This Kerr superposition results in the number state superposition in modes C and D of the lower interferometer, the N00N state.

Figure 8. In the absence of strong Kerr nonlinearities, adding ancillary beam splitters and detectors can generate an effective Kerr. Particular outcomes, such as detecting one and only one photon in the uppermost and lowermost detectors generates an N = 4 N00N state with the number-state inputs $|3\rangle_A |3\rangle_B$ as shown. The first beam splitter produces the combination of number states indicated. The projective measurement collapses these amplitudes into only the $|1\rangle_A |3\rangle_B$ and $|3\rangle_A |1\rangle_B$ components, which upon passage through the second beam splitter produces the N00N state.

Figure 9. It is in fact easy, experimentally, to produce an *N = 2* N00N state using a spontaneous parametric downconverter, which approximately produces the state $|1\rangle_A |1\rangle_B$, as shown. Upon passage through an ordinary 50-50 beam splitter, this state is converted into the N00N state via the Hong-Ou-Mandel effect. Parametric down conversion occurs in special nonlinear crystals when, for example, a single ultraviolet photon is converted into two daughter photons, as show in the inset.

Figure 10. In this plot we reproduce the graph of *M* from figure 6, but now with 50% or 3 dB of loss introduced into one arm of the interferometer. Due to the super-Beer's law, the visibility of the N00N state fringes (green) shrinks much more rapidly than those of the coherent state (red). Note that the maximum slope of the green and red curves is now about the same, indicating they same shot-noise sensitivity. Nevertheless, we can still observe the multiple fringes, and so it is possible in the presence of loss for N00N states to show super resolution without super sensitivity.

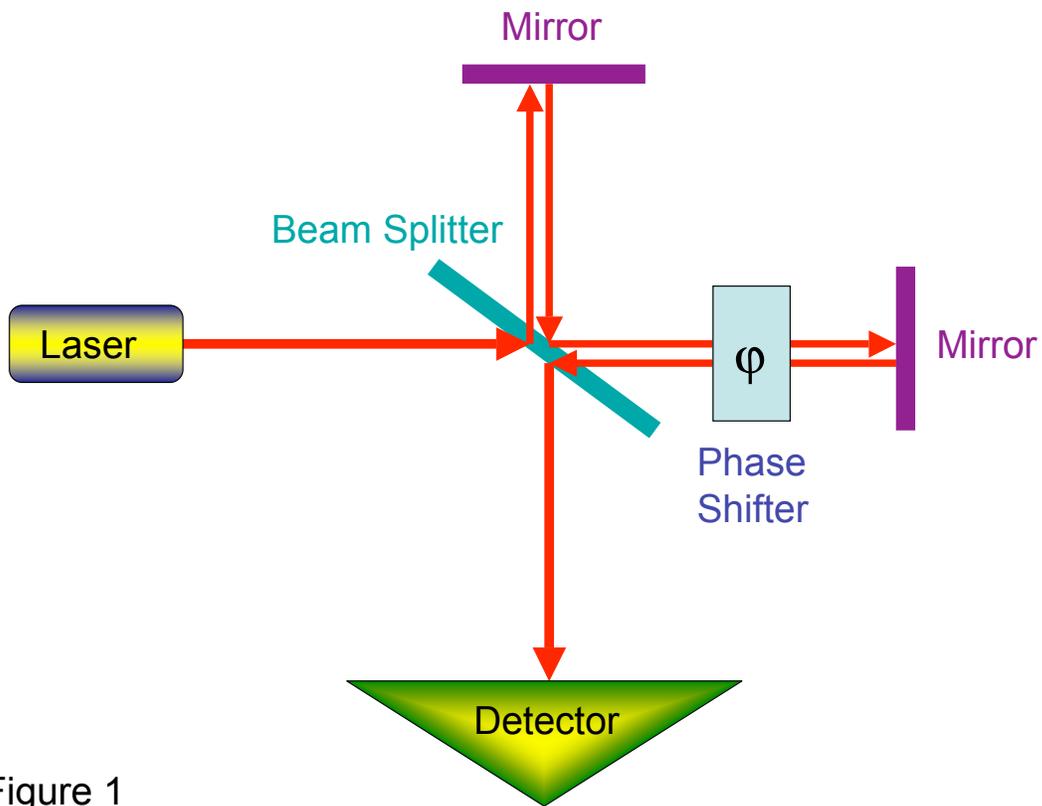

Figure 1

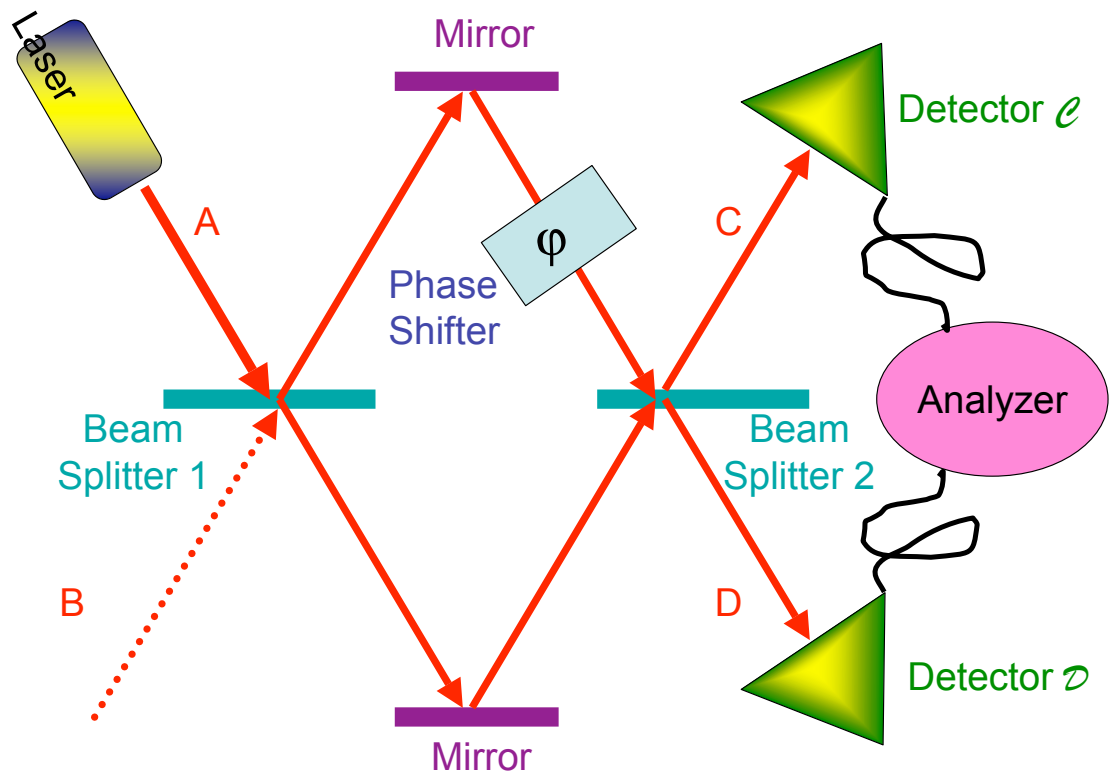

Figure 2

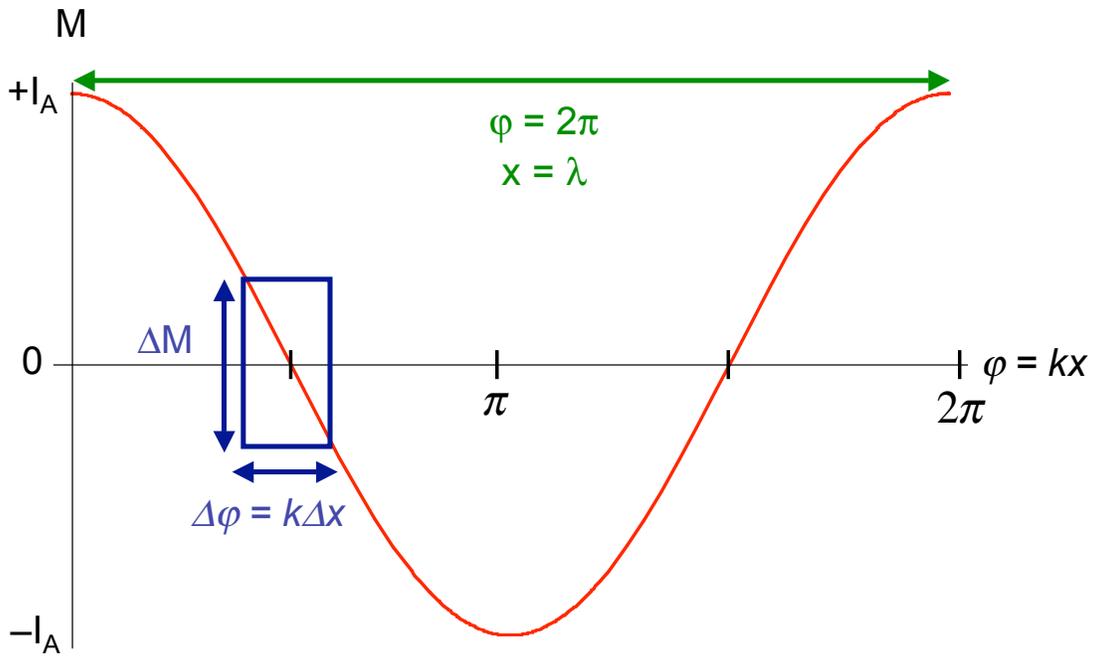

Figure 3

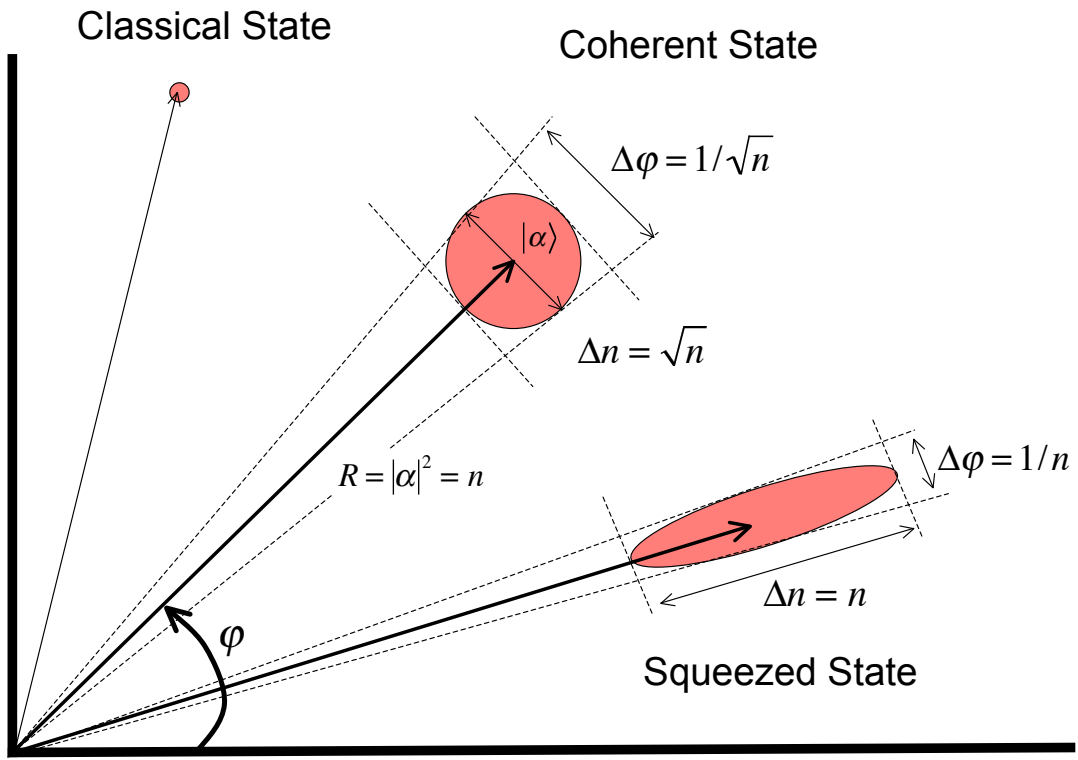

Figure 4

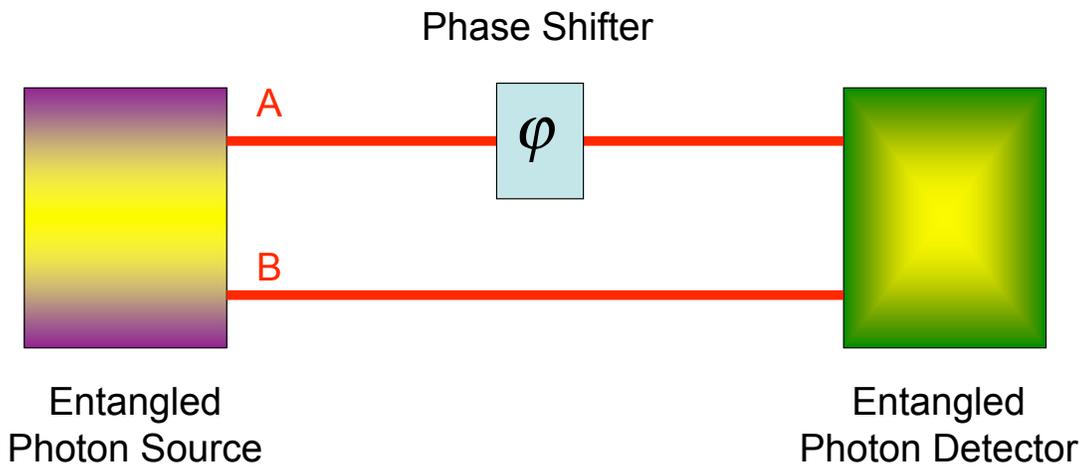

Figure 5

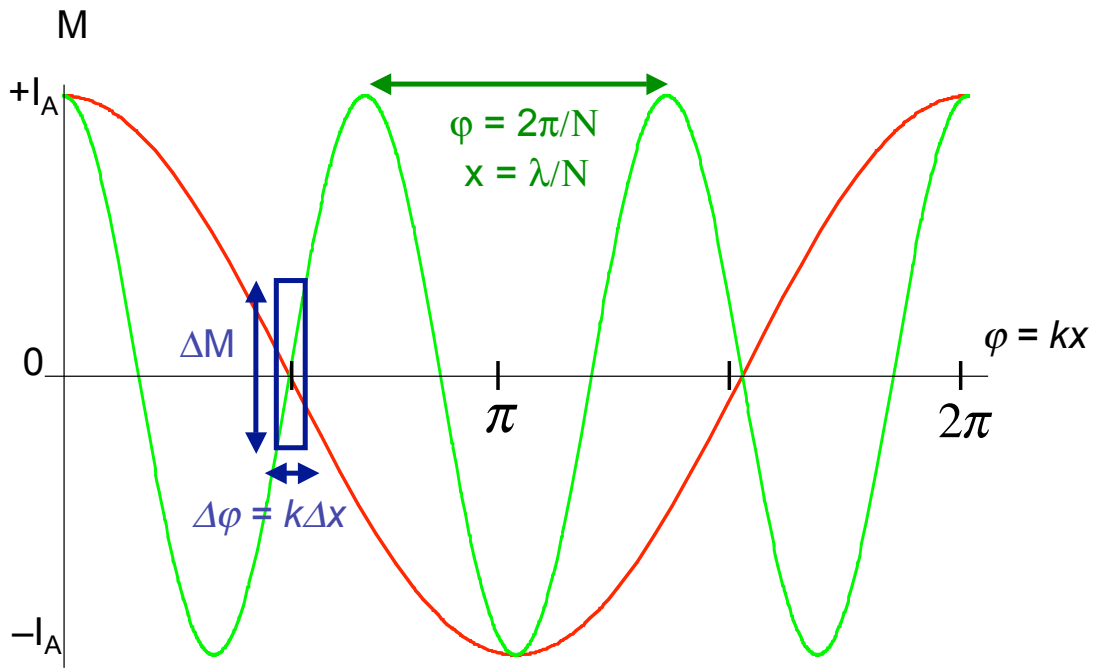

Figure 6

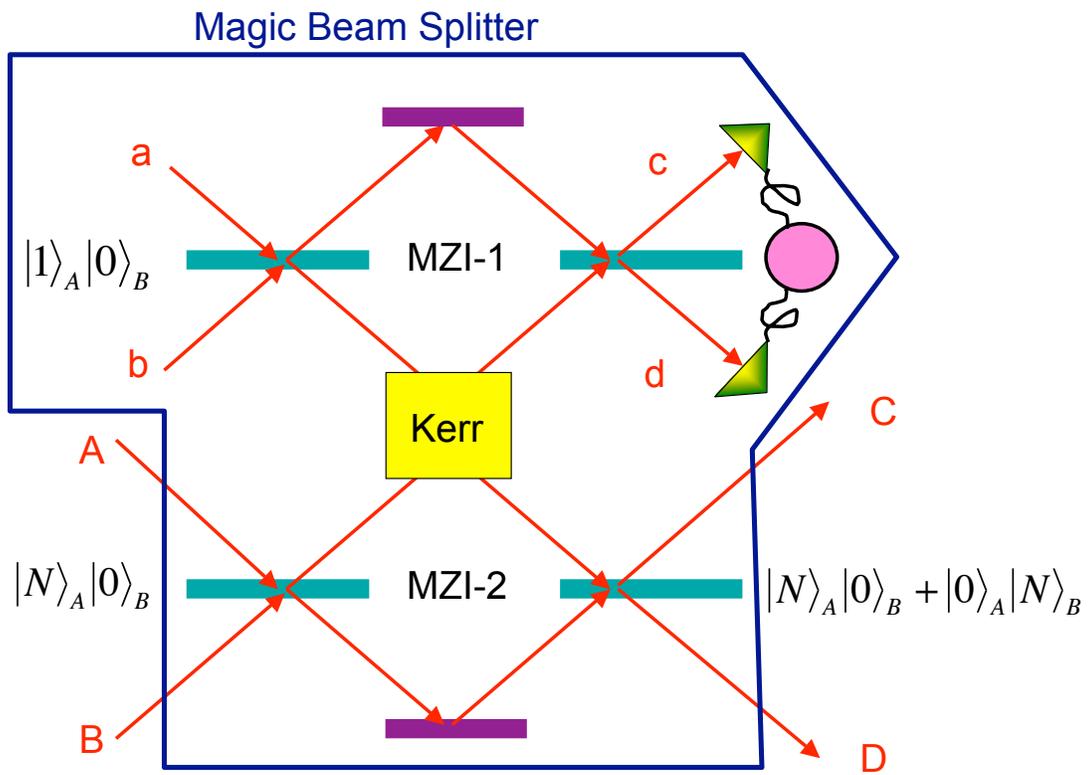

Figure 7

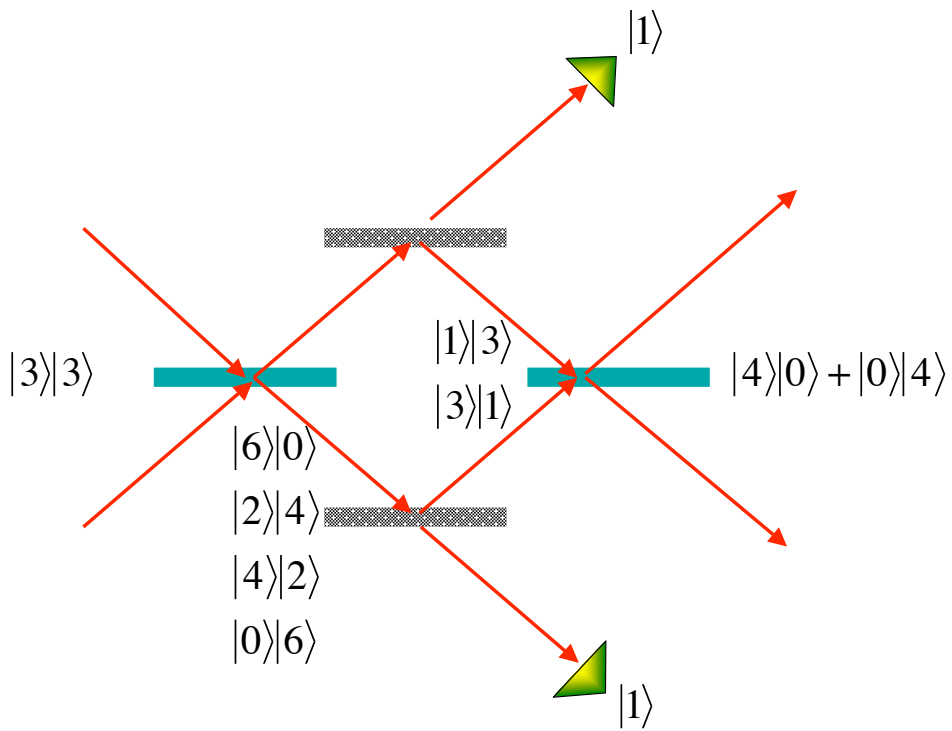

Figure 8

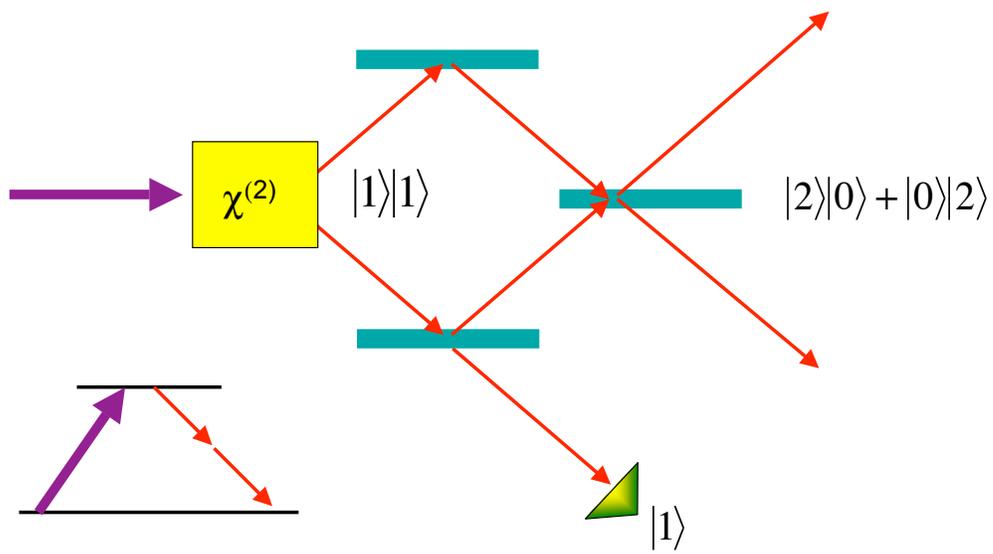

Figure 9

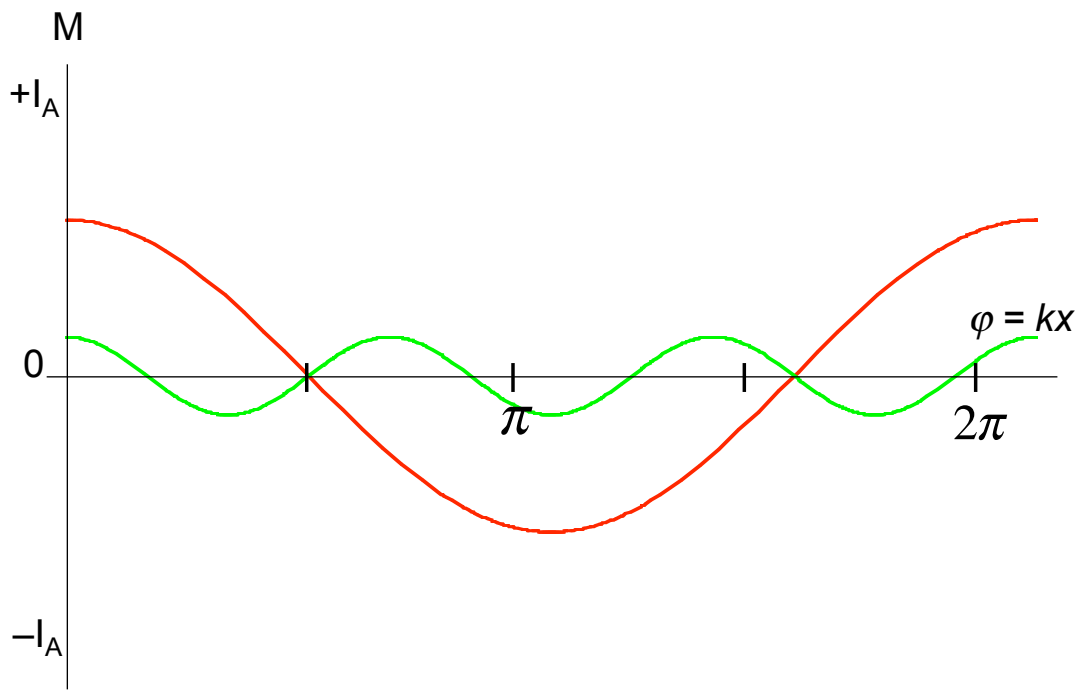

Figure 10